\documentclass[amssymb,useAMS,prd,aps,amsmath,twocolumn,superscriptaddress,nofootinbib,preprintnumbers]{revtex4-1}
\usepackage{graphicx}
\usepackage{xcolor}
\newcommand{\remove}[1]{}
\usepackage{hyperref}
\usepackage{amsmath}

\begin{document}

\title{
Constraints on General Light Mediators from PandaX-II Electron Recoil Data}
\author{Amir N.\ Khan}
\email{amir.khan@mpi-hd.mpg.de}
\affiliation{Max-Planck-Institut f\"{u}r Kernphysik, Postfach 103980, D-69029
Heidelberg, Germany}

\begin{abstract}
\noindent
PandaX-II has analyzed their complete data set of the electron recoil energy spectrum and has confirmed the XENON1T (1-7) keV excess, although the excess was also found compatible with the total background. Treating the background as well known, in which case it provides a good fit to the observed spectrum, one can expect stronger constraints on any new physics model with this data. With this motivation, we derive constraints on the new general vector (V), axial-vector (A), scalar (S) and pseudoscalar (P) interactions if any of them contribute to the neutrino-electron elastic scattering. The derived constraints on the couplings at $90\%$ C.L., respectively, are $g_{V^{^{\prime }}} \lesssim 32 \times 10^{-7}$ for the mediator mass ${\lesssim 10}$ keV, $g_{A^{^{\prime }}} \lesssim 34 \times 10^{-7}$ for mass ${\lesssim 10}$ keV, $g_{S}\lesssim 49 \times 10^{-7}$ for mass ${\lesssim 20}$ keV and $g_{P} \lesssim 67 \times 10^{-7}$ for mass ${\lesssim 30}$ keV.

\end{abstract}

\date{\today}
\pacs{xxxxx}
\maketitle

\section{Introduction}
Recently, PandaX-II has observed an excess in the electron recoil energy
spectrum \cite{Zhou:2020bvf} following a similar observation by XENON1T \cite{Aprile:2020tmw}. The expected background lies within 1$\sigma$ of the experimental error in the $(1-7)$ keV recoil energy region. However, the PandaX-II collaboration has also confirmed a degeneracy between the XENON1T signal and the tritium background. Having a relatively less exposure than XENON1T, they have to put a relatively weaker constraint on solar axions and anomalous neutrino magnetic moment. The possibility of tritium decay as the source of the excess or the possibility of any new physics or the new physics induced tritium decay will be resolved by the near future direct direction experiments with their large exposure and unprecedented low background \cite{Zhang:2018xdp, Aprile:2020vtw,Akerib:2019fml,Aalbers:2016jon,Aalbers:2020gsn}. 

PandaX-II is a 580-kg dual-phase liquid xenon detector based on the detection technique of capture of photons from the prompt scintillation and of the delayed photons from the ionized electrons through photo-multipliers \cite{Zhou:2020bvf}. For the analysis of ref. \cite{Zhou:2020bvf}, they have used the complete data set from Run 9, Run 10, and Run 11 with a total exposure of 100.7 ton-days which is a factor of 3 less than the XENON1T \cite{Aprile:2020tmw}. Up to the statistics, both experiments give almost similar constraints on the solar axion and neutrino magnetic moment. 

After the first observation of the recoil electron excess by XENON1T \cite{Aprile:2020tmw}, there has been a surge of phenomenological papers, either explaining excess by direct dark matter detection or through neutrino nonstandard interactions while several others derived new constraints on model-dependent parameters or on sterile neutrino interactions \cite{Khan:2020vaf,Boehm:2020ltd, AristizabalSierra:2020edu, Okada:2020evk, Lindner:2020kko,Alonso-Alvarez:2020cdv, Chala:2020pbn, Ge:2020jfn, Amaral:2020tga,Benakli:2020vng, Chigusa:2020bgq, Li:2020naa, Baek:2020owl, Gao:2020wfr, Ko:2020gdg, An:2020tcg, McKeen:2020vpf, Bloch:2020uzh, Budnik:2020nwz,Farzan:2020dds,Khruschov:2020cnf,Kim:2020aua, Bally:2020yid, Arcadi:2020zni, Okada:2020evk,Li:2020naa, Shoemaker:2020kji,Khruschov:2020cnf,Shakeri:2020wvk,Takahashi:2020bpq, Takahashi:2020uio}. In continuation to our previous work \cite{Khan:2020vaf}, in which we used the XENON1T data to derive limits on a wide variety of neutrino nonstandard interactions or to explain the excess, here we carry out an analysis for the same neutrino nonstandard interactions but using PandaX-II data.\ Further, we compare our results with those from ref.\ \cite{Khan:2020vaf} and other experiments. To be more specific, here we will investigate the possibility of general neutrino nonstandard interactions that could modify the neutrino-electron elastic scattering at the low energy end of the observed energy spectrum.

Treating the background reported by PandaX-II as standard model expectations, we will derive constraints on the new light vector (spin-1) and scalar (spin-0) gauge boson masses and their couplings to the neutrinos and electrons in general model-independent vector (V), axial-vector (A), scalar (S) and pseudoscalar (P) interactions. Being similar to XENON1T in all aspects, PandaX-II is sensitive to the same new physics at low energy end of the recoil spectrum. At the lower end recoils, where new physics $\propto$ inverse recoil electron kinetic energy and having a good agreement between the expected background and the observed spectrum, PandaX-II data also has a leverage to put stronger or competitive limits on the light mediator masses and their couplings. We would like to remind that these interactions are predicted by a wide variety of models \cite{Fayet:1977yc,Fayet:2020bmb,Boehm:2004uq,Langacker:2008yv}, however, here we adopt the agnostic approach and do not discuss the possible origin of these interactions. An important aspect of the mediators considered here is that they have very low masses and very weak couplings and the spontaneous symmetry breaking to generate their masses is below the electroweak scale. For other phenomenological implications of such interactions, see refs. \cite{Pospelov:2011ha,Pospelov:2012gm,Harnik:2012ni}.

To get a rough estimate, we will first inspect the expected energy spectrum by choosing some bench mark parameter values and compare this with the data for each interaction. Next, we will perform a complete $\chi^{2}$ analysis to explore the full parameter space of the coupling constants and the mediator masses with both two- and one-parameter fitting. We would like to highlight that the two parameter analysis is the standard to present the excluded boundaries but the one parameter fits help to kinematically distinguish these interactions from each other.

After setting up the formal structure in section II, we discuss the analysis details and results in section II. In section IV, we discuss our summary and conclude.

\section{The Expected Events Spectrum}
We present the important expressions and formulas for the new light gauge boson mediating vector, axial-vector, scalar and pseudoscalar interactions contributing to the neutrino-electron elastic scattering process. The standard model differential cross section for the $\nu -e$ scattering is
\begin{widetext}
\begin{equation}
\frac{d\sigma _{\nu _{\alpha }e}}{dE_{r}}=\frac{%
2G_{F}^{2}m_{e}}{\pi } \left[ g_{L}^{2}+g_{R}^{2}\left( 1-\frac{E_{r}}{E_{\nu }}%
\right) ^{2}-g_{L}g_{R}\frac{m_{e}E_{r}}{E_{\nu }^{2}} \right]
\label{cross-section}
\end{equation}
\end{widetext}
where $G_{F}$ is Fermi constant, $m_e$ is mass of electron, $g_{L}=(g_{V}+
g_{A})/2$+1 for $\nu_{e}$, $(g_{V}+
g_{A})/2$ for $\nu_{\mu,\tau}$, $g_{R}=(g_{V}-g_{A})/2$ for $\nu_{e, \mu,\tau}$ , $g_{V}=-1/2+2\sin ^{2}\theta_{W}$, $g_{A}=-1/2$, $E_{\nu}$ is the incoming neutrino energy and $E_{r}$ is the electron recoil energy in the detector. We take $\sin ^{2}\theta_{W} =0.23867\pm 0.00016$ in the $\overline{\text{MS}}$ scheme \cite{Zyla:PDG2020} with small radiative corrections, less than $2\%$, included.

In the following we assume additional model independent light Spin-1 $(Z^{\prime}_{\mu})$ and spin-0 $(S)$ mediators which couple to electrons and to the three flavor of neutrinos with equal coupling strengths via vector, axial-vector, scalar and pseudoscalar interactions. Such interactions are described by the following Lagrangians:

\begin{align}
\mathcal{L}_{V^{\prime}}&=\   
 -g_{V}^{^{\prime}}\left[\overline{\nu }_{L}\gamma
^{\mu }\nu _{L}+\overline{e}\gamma
^{\mu }e\right]Z^{^{\prime}}_{\mu} \ \ \ (Vector),\\[10pt]
\mathcal{L}_{A^{\prime}}&=\  
 -g_{A}^{\prime}\left[\overline{\nu }_{L}\gamma
^{\mu }\gamma_{5}\nu _{L}+\overline{e}\gamma
^{\mu }\gamma _{5}e\right]Z^{^{\prime}}_{\mu } \ (Axialvector), \\[10pt]
\mathcal{L}_{S}&=\ -g_{S}\left[\overline{\nu }_{R}\gamma_{5}\nu_{L}+\overline{e}e\right]S+h.c  \ \ \ \ \ (Scalar),\\[10pt]
\mathcal{L}_{P}&=\ -g_{P}\left[\overline{\nu }_{R}\gamma_{5}\nu
_{L}+i\overline{e}\gamma_{5}e\right]S+h.c \  (Pseudoscalar).
\label{lagrangians}
\end{align}
Here $g_{V}^{^{\prime}}, g_{A}^{^{\prime}}, g_{A}$ and $g_{P}$ are the coupling constants for each interaction. Since the vector and axial-vector interactions interfere with the standard model vector and axial-vector interactions, so in the limit of low momentum transfer, the SM couplings with electrons, $g_{V/A}$ in eq. $(\ref{cross-section})$ can be replaced by the effective parameters $\widetilde{g}_{V/A}$ \citep{Lindner:2018kjo} as
\begin{align}
\widetilde{g}_{V/A} = g_{V/A}+\left( \frac{g_{V^{^{\prime }}/A^{^{\prime }}}^{2}}{\sqrt{2}%
G_{F}(2m_{e}E_{r}+m_{V^{^{\prime }}/A^{^{\prime }}}^2)}\right) ,
\label{gvga}
\end{align}
where $g_{V^{^{\prime }}/A^{^{\prime }}}$ is the coupling constant and $m_{V^{^{\prime}}/A^{^{\prime }}}$ is the mass of the new vector/axial-vector
mediators.

The contribution of scalar mediators is added without interference. In this case, the scalar and pseudo-scalar interaction cross sections  \citep{Cerdeno:2016sfi} are%
\begin{align}
\left( \frac{d\sigma _{\nu _{\alpha }e}}{dE_{r}}\right) _{_{S}}=\left( 
\frac{g_{S}^{4}}{4\pi (2m_{e}E_{r}+m_{S}^2)^{2}}\right) \frac{%
m_{e}^{2}E_{r}}{E_{\nu }^{2}} \label{gs}, \\ \
\left( \frac{d\sigma _{\nu _{\alpha }e}}{dE_{r}}\right) _{_{P}}=\left( 
\frac{g_{P}^{4}}{8\pi (2m_{e}E_{r}+m_{P}^2)^{2}}\right) \frac{%
m_{e}E_{r}^2}{E_{\nu }^{2}},
\label{gp}
\end{align}
where $g_{S}$ and $g_{P}$ are the scalar and pseudoscalar coupling constants and $m_{S }$, $m_{P }$ are, respectively, their masses.

To estimate the new physics contribution to the observed electron recoil spectrum at PandaX-II, we define the differential event rate in terms of the reconstructed recoiled energy $(E_{rec})$ as%
\begin{widetext}
\begin{equation}
\frac{dN}{dE_{rec}} = N_{e}\int_{E_{r}^{th}}^{E_{r}^{mx}}dE_{r} \int_{E_{\nu }^{mn}}^{E_{\nu }^{mx}}dE_{\nu
}\left( \frac{d\sigma _{\nu _{e}e}}{%
dE_{r}}\overline{P}_{ee}^{m} +\cos^2{\theta_{23}} \frac{d\sigma _{\nu _{\mu }e}}{dE_{r}}\overline{P}_{e\mu}^{m} +\sin^2{\theta_{23}} \frac{d\sigma _{\nu _{\tau}e}}{dE_{r}}\overline{P}_{e\tau}^{m}\right) \frac{d\phi }{dE_{\nu }}\epsilon (E_{rec})G(E_{rec},\
E_{r}),
\label{rate}
\end{equation}
\end{widetext}
where $G(\ E_{rec}, E_{r})$ is the Gaussian function which takes into
account the finite energy resolution of the detector with a resolution power $%
\sigma (E_{r})/E_{r}$ and $\epsilon
(E_{rec})$ is the detector efficiency taken from ref.\ \cite{Wang:2020coa},
$\ d\phi /dE_{\nu }$ is the solar flux spectrum taken from ref.\ \cite{Bahcall:2004mz} and
$N_{e}$ is the 1007.7 ton-day exposure of PandaX-II \cite{Zhou:2020bvf}.
Here, $d\sigma _{v_{\alpha }e}/dE_{r}$ are cross sections
given in eqs. $(\ref{cross-section})$, $(\ref{gvga})$, $(\ref{gs})$ and $(\ref{gp})$ above, $\overline{P}_{ee}^{m}$ $\ $ and $\overline{P}_{e\mu /\tau }^{m}$ are the oscillation-length averaged
survival and conversion probabilities of solar neutrino
including the small matter effects as given by 
\begin{equation}
\overline{P}_{ee}^{m}=s_{13}^{4}+\frac{1}{2}c_{13}^{4}{}(1+\cos 2\theta _{12}^{m}\cos
2\theta _{12})
\label{prob}
\end{equation}%
and $\overline{P}_{e\mu /\tau }^{m}=1-\overline{P}_{ee}^{m},$ where s$_{ij}$, c$_{ij}$ are mixing
angles in vacuum and $\theta _{12}^{m}$ is the matter effects induced mixing
angle taken from \cite{Lopes:2013nfa, Zyla:PDG2020}. We take values of oscillation parameters and their uncertainties from \cite{Zyla:PDG2020} and for the analysis we consider only the normal ordering scheme of the neutrino masses. 
The integration limits are $E_{\nu
}^{mn}=(E_{r}+\sqrt{2m_{e}E_{r}+E_{r}^2})/2$ and $E_{\nu }^{mx}$ is the upper limit of each component of the PP-chain and CNO neutrinos which were taken from ref. \cite{Bahcall:2004mz}. Note the CNO neutrino has negligibly small effect for the observed energy range of interest. 
$E_{r}^{th}=1\ $keV is the detector threshold and $E_{r}^{mx}=25$ keV is the maximum recoil energy for the region
of interest. We further note that we do the analysis with the general case of non-maximal scheme of ``23'' sector as clear from eq. $(\ref{rate})$.  
\begin{figure}[tp]
\begin{center}
\includegraphics[width=3.5in, height=5.3in]{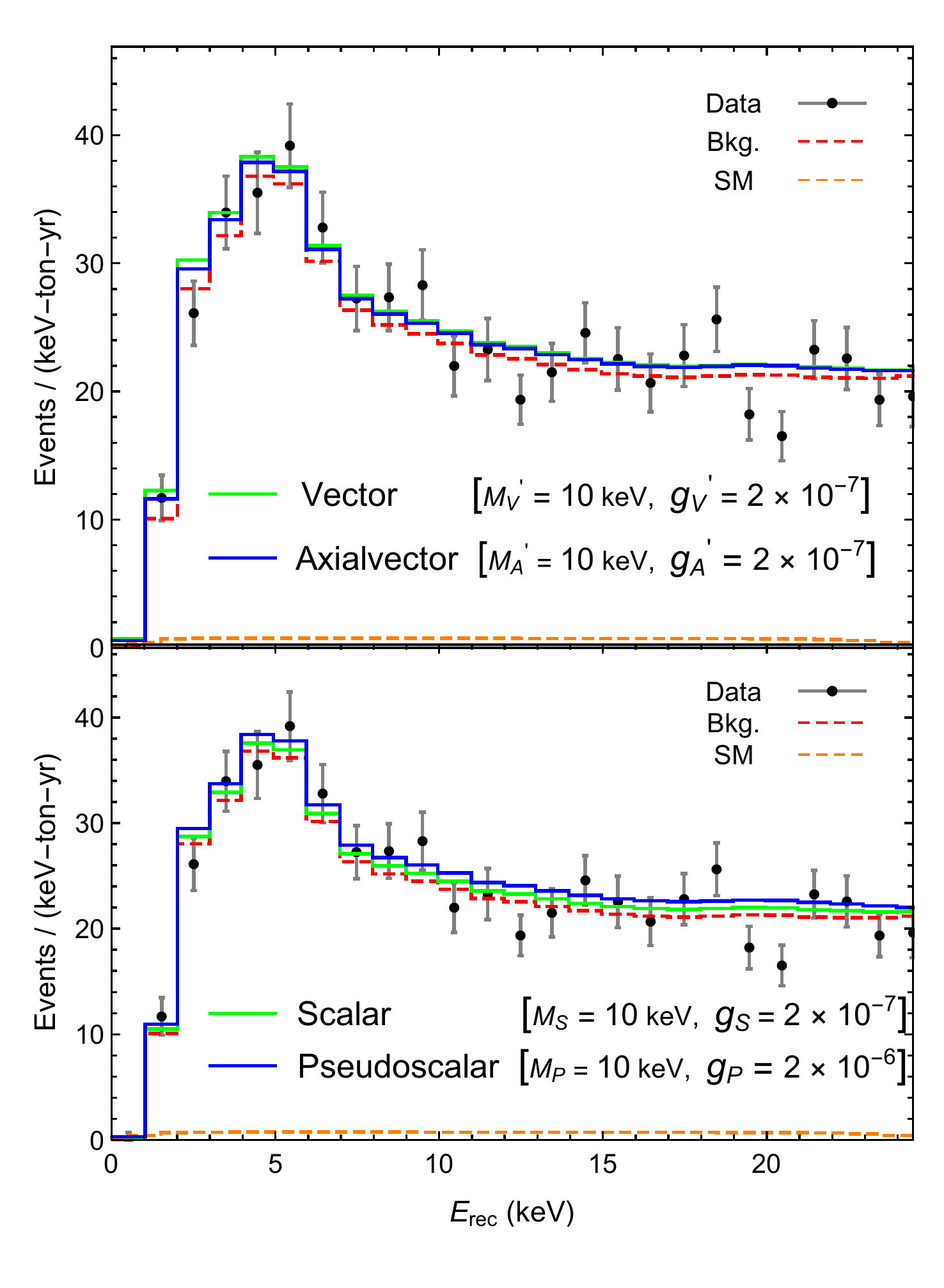}
\label{nuem}
\end{center}
\vspace*{-1cm}
\caption{Experimental data, backgrounds, the standard model expectation and the new physics spectrum with benchmark values of the mediator masses and their couplings for vector/axial-vector $(top)$, scalar/pseudoscalar $(bottom)$. The data points and background were taken from ref. \cite{Zhou:2020bvf}.}
\label{spectrum}
\end{figure}
\section{Analysis and Results}
We take all the data points, their errors and the respective background from ref.\ \cite{Zhou:2020bvf}, as also shown in our fig.\ \ref{spectrum}. The total background shown is the sum of the background from detector source components. They mainly include the Flat ER, Tritium, ${}^{127}Xe$, ${}^{136}Xe$, accidental and from neutrons.  
With the above set up, we calculate the differential event energy spectrum as a function of $E_{rec}$ for the standard model case and for the new interactions considered here.

In addition to the measured background reported in ref.\ \cite{Zhou:2020bvf}, we also calculate the standard model contribution  of solar ${\nu-e}$ scattering to the total expected spectrum as shown with orange color in fig. \ref{spectrum}. As can be seen in the figure, this contribution is although small, but not entirely negligible and becomes significant in interference with the new vector and axial-vector interactions. For the scalar and pseudoscalar, we add new physics as a signal above the given background plus the small standard model contribution as there is no interference with SM in this case. More importantly, we would like to emphasize that our results mainly rely on reproducing the PandaX-II neutrino magnetic moment result since we adjust our $\chi^2$ function to first exactly reproduce the PandaX-II upper limit of $3.2\times 10^{-11}\mu _{B}$ at 90\% C.L.. We discuss this in more detail in the next section.

\subsection{The spectral shape analysis}
We perform a spectral shape analysis of the measured spectrum for each type of interaction. We choose bench mark values of the coupling constants $1\sigma$ experimental error at fix the values of the mediator masses at 10 keV in each case. The results of this exercise are shown in fig.\ \ref{spectrum}(top) for the vector and axial-vector interactions and in fig.\ \ref{spectrum}(bottom) for the scalar and pseudoscalar interactions. Here note that this analysis was done for illustration purposes to get a rough estimate of the parameter range and choices. It is obvious that changing the masses values, the best-fit values of the coupling constants will change accordingly. As shown in fig.\ \ref{spectrum}, within $1\sigma$ experimental error the preferred values are $2\times 10^{-7}$ value for the coupling constants of the vector, axial-vector and scalar interactions while $2\times 10^{-6}$ for the pseudoscalar interactions at the same mass value of $10$ keV in each case.
\subsection{Constraints on V,A, S and P interactions from the PandaX-II data}
\begin{figure}[tp]
\begin{center}
\includegraphics[width=3.3in, height=3.1in]{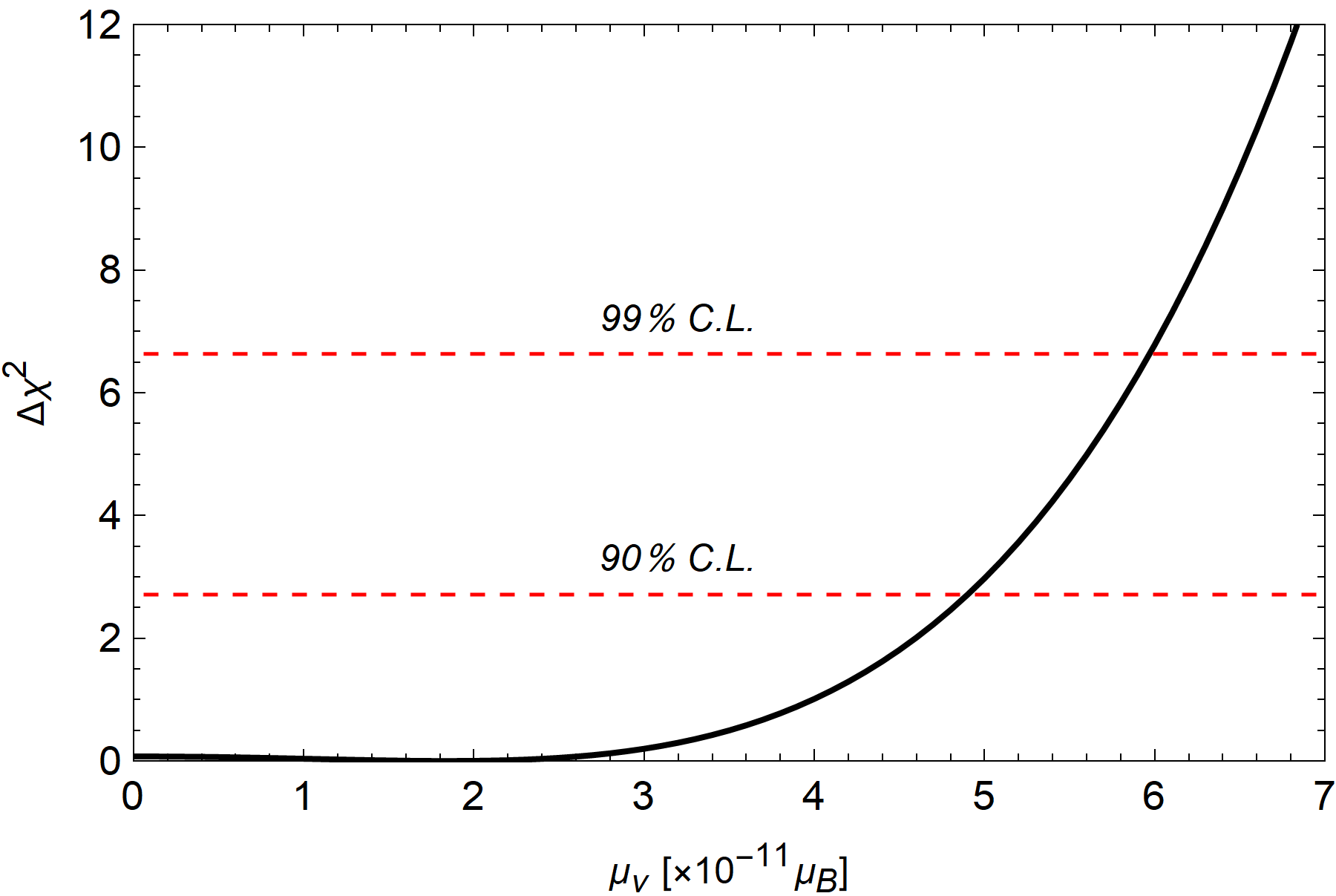}
\label{nuem}
\end{center}
\vspace*{-0.5cm}
\caption{$\Delta \chi^{2}$ distribution of solar neutrino effective magnetic moment from PandaX-II data using eq. $(\ref{chisq})$ with $90\%$ and $99\%$ C.L. projections. The $90\%$ C.L. upper limit exactly corresponds to $4.9 \times 10^{-11} \mu_{B}$ which is the value reported by PandaX-II collaboration.}
\label{NMM}
\end{figure}

\begin{figure*}[tp]
\begin{center}
\includegraphics[width=7.2in,height=5.8in]{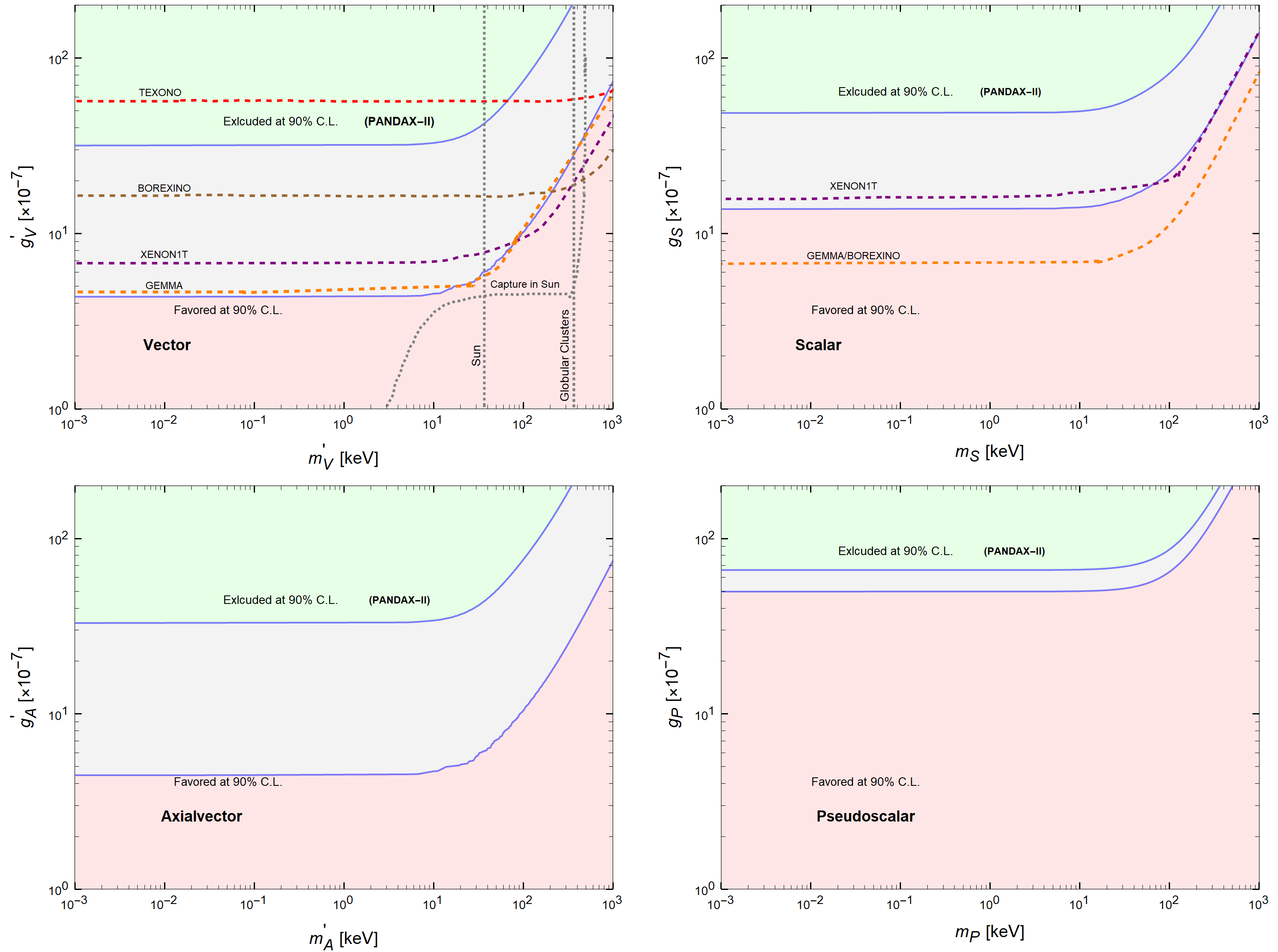}
\end{center}
\caption{The 90\% C.L. allowed regions in the parameters spaces of the light mediator masses and their couplings using PandaX-II data. The green color in each graph represents the excluded region while the red colors correspond to the favored regions at the 90\% C.L. corresponding to the two degrees of freedom. The grey area in each plot is the overall included region. The overlaid curves for the other terrestrial experiments (dashed lines), GEMMA  \cite{Beda:2009kx}, Borexino \cite{Agostini:2018uly}, TEXONO \cite{Deniz:2010mp}, XENON1T \cite{Aprile:2020tmw} and for the astrophysical data (dotted line) were taken from refs. \cite{AristizabalSierra:2020edu, Boehm:2020ltd, Harnik:2012ni} are shown in.}
\label{2d_mediators}
\end{figure*}

\begin{figure*}[tp]
\begin{center}
\includegraphics[width=7.2in, height=5.3in]{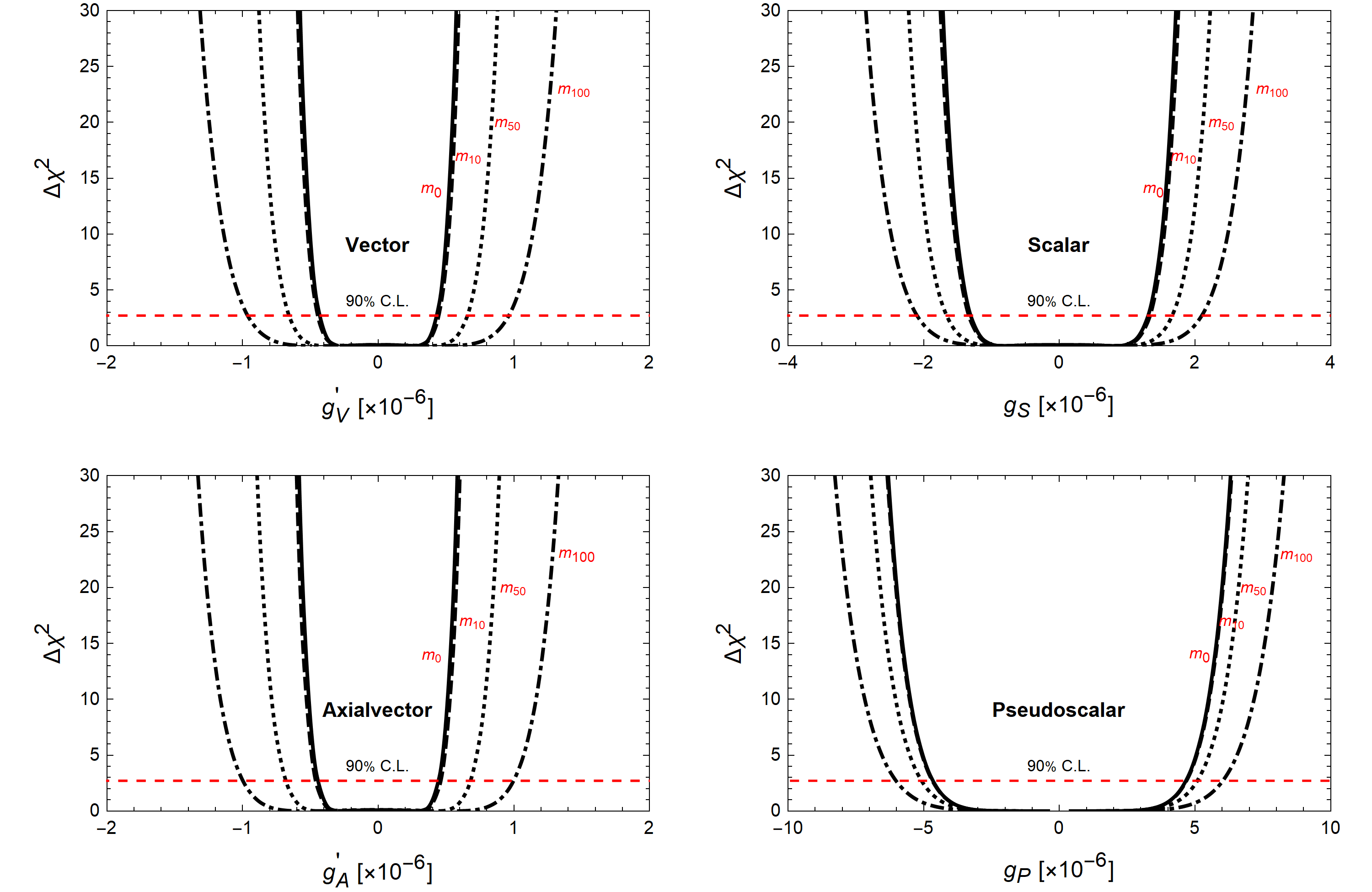}
\label{nuem}
\end{center}
\vspace*{-1cm}
\caption{{}\textbf{\ }1 parameter $\Delta\chi^2$ distributions and the 90\% C.L. projection of vector, axial-vector, scalar and pseudoscalar interaction couplings for the benchmark values of the corresponding mediator masses $(m_{0}$= 0 keV, $m_{10}$= 10 keV, $m_{0}$= 50 keV and $m_{100}$= 100 keV). Notice the difference in scales on x-axes for the scalar/pseudoscalar and vector/axial-vector cases.}
\label{1d}
\end{figure*}

To derive constraints on the coupling constants and the mediator masses of the new interactions considered using the PandaX-II data, we define a modified $\chi^{2}$ function as follows 
\begin{equation}
\chi ^{2}=\underset{i,j}{\sum }a\left( \frac{%
b_{i}(dN/dE_{rec}+B(E_{rec}))_{th}^{j}-(dN/dE_{rec})_{obs}^{j}}{c_{i}\sigma
^{j}}\right)^2, 
\label{chisq}
\end{equation}
where $a$, $b_{i}$ and $c_{i}$ with $i=1,2$ are the scaling factors introduced here in order to reproduce the upper bound on the enhanced neutrino magnetic moment as obtained by PandaX-II in ref. \cite{Zhou:2020bvf}. These factors have the values $%
a=10^{-4},b_{1}=1,\ b_{2}=0.11, c_{1}=0.15,\ c_{2}=1.$ In actual, these scaling factors
account for the uncertainties related to the background and for other systematic errors in the efficiencies, detector mass, etc \cite{Zhou:2020bvf}. The expression in the bracket
(.....)$_{th}^{j}$ corresponds to the expected number of events in the $j-$th bin
and the observed numbers of events are represented by the terms in the bracket (...)$_{obs}^{j}$. 
$\sigma ^{j}$ is the total uncertainty in the corresponding $j$-th bin which were obtained from the data error bars of fig.\ \ref{spectrum} in ref.\ \cite{Zhou:2020bvf}. Using this $\chi ^{2}$ function of eq. $(\ref{chisq})$, we can exactly reproduce the upper bound of $4.9 \times 10^{-11}\mu _{B}$ at 90\% C.L. \cite{Zhou:2020bvf} on the neutrino magnetic moment as shown in fig. \ref{NMM}. For our analysis, we take a total of 25 energy bins.
Using the fitting function in eq.\ $(\ref{chisq})$ and the setup developed above, we now proceed to explore the two-parameter allowed region and the one-dimensional $\Delta \chi^{2}$ distribution of coupling constants at fixed values of masses for all types of new interactions considered in this work. In the case of the two-parameter fitting, we fit mass and the coupling constant for each interaction with $\Delta \chi^{2}$ corresponding to the two degrees of freedom. The results of this analysis at 90\% C.L. are shown in fig. $\ref{2d_mediators}$ along with the overlaid curves from the other experiments in the case of vector and scalar interactions. For one parameter fits, we show the $\Delta \chi^{2}$ distribution of the coupling constants with four benchmark values of the mediator masses (0 keV, 10 keV, 50 keV and 100 keV). The results obtained from this analysis with 90\% C.L. projections are shown in fig.\ $\ref{1d}$.

\begin{table*}[t]
\begin{center}
\begin{tabular}{c|c|c|c|c|c}
\hline \hline
Coupling & PandaX-II (\textbf{this work}) & GEMMA & XENON1T & Borexino & TEXONO \\ \hline
$g_{V^{^{\prime }}}(\times 10^{-7})$ & $\lesssim 32$ & $\lesssim 5.0$ & $\lesssim 7.0$ & $\lesssim 17$ & $\lesssim 58$ \\ \hline
$g_{A^{^{\prime }}}(\times 10^{-7})$ & $\lesssim 34$ & $-$ & $-$ & $-$ & $-$ \\ \hline
$g_{S}\ (\times 10^{-7})$ & $\lesssim 49$ & $\lesssim 6.0$ & $\lesssim 16$ & $\lesssim 6.0$ & $-$ \\ \hline
$g_{P}\ (\times 10^{-7})$ & $\lesssim 67$ & $-$ & $-$ & $-$ & $-$ \\ \hline \hline
\end{tabular}%
\end{center}
\caption{90\% C.L. (2 dof) upper bounds on the coupling constants of four types of interactions considered in this work. These limits correspond to the mediator masses: $m_{V}^{\prime}/m_{A}^{\prime}$ ${\lesssim 10}$ keV for the vector and axial-vector interactions, ${m_{S}\lesssim 20}$ keV for the scalar and ${m_{P}\lesssim 30}$ keV for the pseudoscalar interactions. These bounds can also be read off directly from fig. \ref{2d_mediators}.}
\label{tab:bounds}
\end{table*}

\section{Results and Discussion}
All results of this study  are shown in fig.\ \ref{2d_mediators} and \ref{1d}. As can be seen in fig.\ \ref{2d_mediators}, in each case the coupling grows for the mediator masses, $m_{V}^{\prime}/m_{A}^{\prime}\gtrsim 10$ keV, for the vector and axial-vector interactions and for $m_{S}\gtrsim 20$ keV for scalar and $m_{P}\gtrsim 30$ keV for the pseudocalar interactions. This behavior is more clearly visible in fig. \ref{1d} where the 1-dimensional $\Delta \chi^{2}$ curves of the coupling strengths are indistinguishable for the cases of 0 keV and 10 keV mediator masses, but they grow above $10$ keV in all cases. Especially, for the pseudoscalar interactions, one can see that a smaller deviation occurs for the larger masses above $\sim 10$ keV.

On the other hand, as can be seen in fig.\ \ref{2d_mediators}, below $\sim 10$ keV mediator masses, the limits on the coupling constants from PandaX-II data extends up to  $32 \times 10^{-7}$ for vector interactions and $34 \times 10^{-7}$ for axial-vector interactions while below $\sim 20$ keV mediator masses, $49 \times 10^{-7}$ for the scalar interactions and below $\sim 30$ keV mediator masses, $67 \times 10^{-7}$ for the pseudoscalar interactions.
In fig.\ \ref{2d_mediators}, in each plot, the regions in red and grey color are, respectively, the favored and included regions while color in green shows the excluded region by PandaX-II data. All the favored and excluded regions were obtained at 90\% C.L. with 2 degrees of freedom. The favored regions in red color are unbounded from below and from the right side for the shown mass ranges. In comparison to this, the XENON1T data can give a limit at one sigma and favored 2$\sigma$ bounded region \cite{Bloch:2020uzh, AristizabalSierra:2020edu, Khan:2020vaf}. The reason is that XENON1T without tritium background is a signal for any new physics at $\gtrsim$ 3$\sigma$. In the case of PandaX-II, in the presence of tritium background, only the excluded limits are possible which are what we have obtained in fig.\ \ref{2d_mediators}.

In fig.\ \ref{2d_mediators}, we also compare the constraints from PandaX-II data for the vector and scalar interactions with other laboratory experiments and with the astrophysics data \cite{Chang:2018rso, Hardy:2016kme}. As can be seen in the figure, for the vector interactions, in the mass range $\lesssim 10$ keV  GEMMA \cite{Beda:2009kx} yields the strongest constraints among all the laboratory experiments. The next stronger constraint comes from XENON1T \ \cite{Aprile:2020tmw,Bloch:2020uzh, AristizabalSierra:2020edu, Khan:2020vaf}, next is Borexino \cite{Agostini:2018uly}, then is PandaX-II (this work) and TEXONO \cite{Bilmis:2015lja} is the weakest of all. The constraints obtained from this work on all types of interactions are summarized in Table\ \ref{tab:bounds} along with the constraints from other laboratory experiments.

On the other hand, the astrophysical constraints are in severe tension with the terrestrial constraints as can be seen in fig.\ \ref{2d_mediators} \cite{Grifols:1986fc,Grifols:1988fv,Farzan:2002wx,Dent:2012mx,Redondo:2013lna,Hardy:2016kme, Heurtier:2016otg,Chang:2018rso,Venzor:2020ova,OHare:2020wah}. However, here is a caveat that helps the terrestrial constraints to evade the astrophysical bounds if taken into consideration.\ The coupling constants and mediator masses have effective dependence upon the environmental conditions of the astrophysical objects such as the matter density and the surrounding temperature. While the conditions for the terrestrial experiments are very different. In particular, for the vector interactions, these limits can be relaxed by several orders of magnitude under the same conditions as shown in ref. \cite{An:2013yfc}. Even the different astrophysical observations could be in tension with each other. For instance, given the very tighter constraints on the coupling and the mass parameters, in the core- collapse supernovae, the $\nu-e$ interactions via the light mediator may also lead to $\nu-\nu$  and $e-e$ interactions which in turn have a reduced energy loss and the production of fewer neutrinos than expected from the astrophysical objects. On the other hand, for the same couplings, the energy transport in the red giants, sun, globular cluster stars are larger than expected, resulting in faster burning of helium fuel which would change the shape of three objects than the observed ones. Likewise, these bounds are also in tension with those from big bang nucleosynthesis. The relatively larger couplings of the terrestrial experiments imply a different thermal history of the universe. This results in an increase in the effective number of neutrinos in the early universe.  \cite{Masso:1994ww,Redondo:2013lna,Ahlgren:2013wba,Hardy:2016kme,Kamada:2018zxi,Escudero:2019gzq,Dutta:2020jsy,Luo:2020fdt}. These bounds can also be circumvented using the same argument as in the case of the astrophysical bounds discussed above.
\section{Summary and Conclusions}
Following the XENON1T observation of the excess at the low energy end of the recoil electrons, PandaX-II has also analyzed their complete data set with the total exposure of 100.7 ton-days \cite{Zhou:2020bvf}. The observed spectrum at the low energy end, (1-7) keV, shows similar behaviour to that of XENON1T, although, PandaX-II collaboration has confirmed that this excess is also consistent with both the total background including tritium and to the XENON1T signal. With no definite conclusion they have constrained the solar axion-electron coupling and the enhanced neutrino magnetic moment.

Using the PandaX-II data here we have constrained the new light vector (spin-1) and scalar (spin-0) mediator masses and their coupling constants for the vector, axial-vector, scalar and pseudoscalar interactions if any of them contributes to the $\nu-e$ elastic scattering process. Such interactions and the light mediators are predicted by a wide class of models in various mass ranges of the mediators, heavy, intermediate, light and very light and weakly coupled. One interesting aspect of such models is that the mass generation of the light mediators occurs at a scale below the electroweak scale. Here, we have considered only their general model-independent form. Further, for simplicity, we have considered similar couplings for electrons and all flavors of neutrinos in each interactions. 

With this setup, we have fitted the PandaX-II spectrum above the total background including the tritium to our expected event energy spectrum for the four types of new interactions. After estimating the parameters ranges by spectral analysis we introduced a modified $\chi^{2}$ function to reproduce the PandaX-II result of the neutrino magnetic moment which is shown in fig.\ \ref{NMM}. This $\chi^{2}$ function was further used to derive constraints on the new light mediator masses and their couplings. 
As mentioned before, our analysis strictly relies on the analysis of ref.\ \cite{Zhou:2020bvf} which already have derived competitive limits on the solar axion couplings and the neutrino magnetic moment. The competitive results of PandaX-II are due to a better agreement between the standard expected background and the data. This agreement leaves a narrower room for any new physics. In return, one can expect better or competitive constraints on any new physics with the current PandaX-II data. In fig.\ \ref{2d_mediators}, one can compare the excluded regions of this study with those from the other terrestrial experiments and astrophysical data for the vector and scalar interactions. The constraints on the coupling constants are also summarized in Table \ref{tab:bounds} for definite ranges of the mediator masses. One can see that PandaX-II with the current data does only better than TEXONO. Particularly, for a fair comparison with XENON1T, both being direct dark matter detection experiments, one should note that XENON1T has 3 times larger exposure than PandaX-II. This fact is consistently reflected in the obtained bounds given in Table \ref{tab:bounds}. Constraints from XENON1T on the coupling constants are stronger only by factor 4.5 than PandaX-II for the vector interactions while factor of 3 stronger for the scalar interactions. 

In conclusion, PandaX-II adds one more candidate to the list of experiments that provides competitive limits on the light mediator masses and their couplings in interactions to neutrino and electrons. Pertinent to the question of the observed excess at both PandaX-II and XENON1T, neutrinos could be the inevitably strong candidate to explain this if the tritium background is excluded in future upgrades of the two experiments or by other near future similar experiments \cite{Zhang:2018xdp, Aprile:2020vtw,Akerib:2019fml,Aalbers:2016jon,Aalbers:2020gsn} or by the dedicated neutrino experiments in the same energy range. In the meantime, limits on the mediator masses and effective couplings of the new vector, axial-vector, scalar and pseudoscalar interactions with neutrinos and electrons are getting improved further. Being the dark dark matter experiment with similar observation as of XENON1T, the PanadaX-II is expected to play a complementary role in both direct dark matter detection and in the neutrino nonstandard interactions.

\begin{acknowledgments}
I thank Ke Han and Xun Chen for providing useful information about PandaX-II. This work is financially supported by Alexander von Humboldt Foundation under the postdoctoral fellowship program.
\end{acknowledgments}

\bibliography{biblio}

\end{document}